# Draw out Carbon Nanotube from Liquid Carbon

Shuang Zhang[1,2,3] , Takeo Hoshi[1,3], Takeo Fujiwara[1,3]

[1]*Core Research for Evolutional Science and Technology (CREST-JST), Japan Science and Technology Agency, 4-1-8 Honcho, Kawaguchi-shi, Saitama 332-0012, Japan*

[2]*Key Laboratory on the Physics and Chemistry of Nanodevices and Department of Electronics, Peking University, Beijing 100871, China*

[3]*Department of Applied Physics, University of Tokyo, 7-3-1 Hongo, Bunkyo-ku, Tokyo 113-8656, Japan*

**Carbon nanotube (CNT) is expected for much more important and broader applications in the future, because of its amazing electrical and mechanical properties[1]. However, today, the prospect is detained by the fact that the growth of CNTs cannot be well controlled. In particular, controlling the chirality of CNTs seems formidable to any existing growth method. In addition, a systematic method for a designed interconnected network has not been established yet, which is focused particularly in nano-electronics field[2-5]. Up to now, there is a strong need for novel synthetic method that is controllable for chirality and architecture and applicable to mass production. Recently, the experimental evidences were reported for the growth of pure carbon-arc-product nanotubes from a liquid precursor[6]. Here we show the simulation results of elongated growth process of CNT by drawing out a well-formed nanotube from liquid carbon and discuss the mechanism and suitable synthesis conditions. The process is also simulated for creation of a Y junction from two isolated CNTs as first step to CNT network. We propose this novel synthetic method as a promising one for catalyst-free mechanical design of interconnected CNT network and CNT cloning.**

The beads of amorphous carbon were found on multi-wall CNTs at the surfaces of nanotube-containing columns as composition of cathode deposit in a pure carbon arc[6]. The beads are considered to be formed at temperatures well below the carbon melting temperature of about 4800 K[7] at the end of a quick cooling process of liquid carbon in a helium atmosphere. The usual regular spacing and similar diameter of the beads are determined by Rayleigh instability. Some of their experimental results imply that, before the formation of the beads, CNTs are crystallized inside liquid carbon at higher temperatures and higher densities than those in the bead formation. Although the detailed atomic picture of nucleation and growth of CNT in cooling liquid carbon has not been clear yet, the experimental evidences of coexistence of CNTs and liquid carbon in cooling motivate us to propose a novel controllable synthetic method of CNTs. Instead the natural nucleating process inside liquid carbon, we use well-formed CNTs to drive the growth process and the morphology of CNTs can be designed by using several controllable conditions.

We present in this letter results of drawing out CNT from liquid carbon by molecular dynamics simulation with electronic structure. As initial configuration, a CNT is put above the liquid surface. The CNT firstly moves down till it plunges into liquid about one nanometer depth, and then turns to rise with a constant velocity. Tens of trials of such process are achieved for showing a growth mechanism and suitable range of synthesis conditions. The present research gives actual atomistic picture of CNT growth at the interface region between liquid carbon and helium atmosphere, which support the scenario proposed by Ref. 6.

Before the simulation of CNT growth, we simulate the cooling process of pure liquid state with a 216-atom periodic cell. The typical density of liquid is 1.7 g/cm$^3$, which corresponds to a pressure on the order of GPa. As a result, the diffusive motion of atoms in such a low density liquid is observed even below the melting temperature (at 4000 K, for instance). In high density liquid (2.5 g/cm$^3$ or above), however, the diffusive motion is suppressed and graphite-like structures appear at the same temperature. The above observation implies that CNTs and liquid carbon in cooling can coexist near the surface in the low density region, which accord with the idea that the observed beads are formed from viscous liquid coating CNTs[6].

Most of our CNT growth simulations were carried out with a 3 nm-long single-wall CNT of the chirality index (9,0). One of successful examples is shown in Fig. 1 (see Supplementary Video 1), where CNT is drawn from liquid of 1.7±0.1 g/cm$^3$ with speed of 30 m/s. The CNT is elongated by longer than 2 nm, and the growth process still continues. The lower cap of CNT melts completely after plunged into liquid for no more than 2 ps. During the growth process, it can be seen that CNT is formed while carbon atoms are just drawn away from liquid surface. The evaporated carbon clusters do not participate in the growth process evidently, but form sometimes chain-like structures. Also we observed that the CNT fluctuates parallel to the liquid surface with a speed of the order of 10 m/s but this movement does not interrupt the growth process. Defects in the elongated part of CNT appear mainly as pairs of five- and seven-membered rings. Therefore, the chirality of the host CNT is not kept strictly with such a quick growth velocity. The diameter of drawn-out CNT is nearly uniform.

Here, we discuss two crucial factors for growth of CNT. One factor is cooling process. In experiment, the local temperature of CNT can be controlled to be lower than that of liquid part by the high-pressure helium atmosphere and other cooling source (e.g. the tip of electron microscope) contacting with CNT. The local temperature distributions of CNT and liquid carbon (Fig. 1b) show that the gradient of temperature decrease near the surface of liquid is much more steep than other parts, mainly because the point contact makes the heat transfer between CNT and liquid is not as smooth as between different parts of the CNT or liquid. This means that the atoms condensing into CNT actually go through a cooling process. In Fig. 1b, the local temperature of the upper CNT tip is set to be 3800 K, which is lower, by 1000 K, than the averaged temperature of the whole system. The growth process is still successful when the temperature difference decreases to 500 K, but cannot proceed while decreasing to nearly zero. Therefore, the cooling process near the surface is crucial for growth of CNT.

The second factor is the electronic structure, particularly, the p bond. The electronic density of state in the low-density liquid carbon has a narrow p band around the Fermi level with the band width of about 10 eV, just as in CNT. Unlike in CNT, however, the p band in liquid state is not clearly split into the bonding and anti-bonding states and non-bonding state appear near the Fermi level. The calculated local density of states shows that non-bonding p electrons are transformed into p-bonding states at the elongated CNT part during the process of elongated CNT growth, which is important for forming the geometry of CNT.

The validation of the present analysis was confirmed in the simulation of cooling process of liquid carbon of 1.7 g/cm$^3$ with the cooling rate similar to that in Fig. 1b, in which graphite-like sheets structures were formed. It was also observed that, when the cooling rate is reduced by a factor 1/5, sheet formation temperature increases by about 600 K. This observation of pure liquid system implies that a growth process with a slower draw-up velocity should be also successful with a smaller temperature difference between the CNT upper tip and the liquid part, because cooling rate is proportional to draw-up velocity. Unfortunately, establishing conclusion by simulations with such slow draw-up velocity is impractical by the present computational resources.

Tens of trials at different liquid densities, temperatures and draw-up velocities are done for understanding relations between the three parameters. The results are summarized in Fig. 2. The temperature conditions of all trials are near melting curve in phase diagram suggested by Bundy[8]. In Fig. 2a, the temperature of the 'whole system' in the simulation should be regarded as the temperature of the interface region in experimental situation, since the present simulation is that of liquid surface region within nanometer scale. From Fig. 2a, CNT grows longer than 1.5 nm within the temperature range from 4600 K to 5000 K. At lower temperatures, we observe that the carbon atoms can not be supplied enough from liquid in the contact point between the CNT and liquid due to the smaller diffusion constant. This means that a slower draw-up velocity is required for the successful growth at these temperatures. On the other hand, elevating temperatures will destroy the CNT growth. Figure 2b shows the results of different densities at the temperature of 4800 K and shows that the successful range of liquid densities is from 1.55 g/cm$^3$ to 1.9 g/cm$^3$. In lower density (< 1.55 g/cm$^3$), the atoms at liquid surface are too sparse to condensate into the CNT efficiently. In higher density (> 1.9 g/cm$^3$), on the other hand, the supply of atoms from liquid surface is not enough due to the fact that the diffusion constant in denser liquid state is smaller.

We also simulate the drawing process of CNT (5,5), (14,6) and double-wall CNT (9,0)/(14,6) with speed of 50 m/s from liquid of 1.7 g/cm$^3$ at 4700 K, and CNTs can grow longer than 1.5 nm. We note that, in almost all cases, the diameter of nanotube keeps nearly uniform in the elongated part, in particular, the diameter of elongated double-wall CNT is almost the same as that of the host CNT, probably because inner and outer wall spatially confine each other. However, we could not try many times enough to establish several conclusions in the optimal synthesis condition for uniformity of the structure of elongated CNT.

We also simulated the creation and growth process of a Y junction drived by two isolated single-wall CNTs (9,0) on the surface of liquid carbon at 4700 K (see Supplementary Video 2). The result is shown in Fig. 3. Two CNTs fall down along the z direction and plunge into liquid carbon of 1.7 g/cm$^3$ one after the other, and then rise simultaneous with the same velocity of 20 m/s. The lower caps of the two initial CNTs melt and a Y junction is formed. The lower part of the Y junctions is created and elongated by longer than 1.5 nm from liquid. The lower part is open-ended and will be elongated further. The similar fabrication processes with the drawn-up velocity of 20-30 m/s and at the temperatures of 4700-4800 K are simulated for three times, and all are successful. More quantitative discussion on the growth conditions should be investigated in future. We can say that the growth mechanism of a Y junction is the same as in the elongation of CNT and the optimal conditions of the two situations should be similar.

In this letter, we propose a novel catalyst-free synthetic method of carbon nanotubes: We plunge carbon nanotubes into liquid carbon with cooling in high-pressure helium atmosphere and draw out with a considerable velocity. The results of simulations show that carbon nanotubes can be elongated steadily. The present discussion implies that a real experimental situation, with a much slower draw-up velocities, could achieve the CNT growth more successful than in the present simulations. The proposed method has two features: One is that growth process including track and velocity of growth can be mechanically controlled. The successful simulations of creation of a Y junction from two nanotubes indicate that the method has advantage of possibilities of mechanical design, which could be applied to large yield growth of interconnected carbon nanotube network. The other feature is that a well-formed nanotube is used to drive the growth process at high temperatures similar to those in the arc-discharge method. Since CNTs produced by arc-discharge are almost defect-free, the high crystallinity of produced CNT can be expected in the present method. Therefore, the present synthetic method is worthy of further experimental research as an attempt to realize clone of carbon nanotubes.

**Methods**

The present calculation was carried out by an order-N electronic-structure theory, in which the Krylov subspace method is used to calculate the one-body density matrix[9] instead of eigenstates and the computational cost is proportional to the system size (N). We use a Slater-Koster type Hamiltonian with Xu's parameters[10]. The computation was carried out mainly using eight Opteron processors of 2.2 GHz. A typical computational time is ten seconds per molecular dynamical step for a system of 2000 atoms. There are already several successful studies of carbon systems using order-N methods with the present Hamiltonian (See recent reviews (11,12)). We confirmed the accuracy of the present order-N method in the simulation of equilibrium liquid carbon, which shows an excellent agreement, between the present calculation and the eigenstate calculation, in the results of radius distribution function and diffusion constant.

In the present CNT growth simulation, the liquid surface lies at z=0 and the simulation cell is periodic in x-y plane with cell parameter of 3 nm × 3 nm, and the length of the z-direction is 15 nm. In some of calculations, the cell of 4 nm × 4 nm is used. The number of atoms per cell is more than 2200. The depth of a liquid part is about 2 nm. The bottom of liquid is set to rigid wall of perfectly elastic collision. In the experimental situation, the evaporated atoms can run away from the surface region, since there is no boundary. In the present simulations, the atoms above the liquid surface by 5 nm are 'cut and pasted' into the liquid region, so as to preserve the number of atoms in the liquid. The time step of all simulations is 1 fs. The Nosé-Hoover thermostat is used to control the total kinetic energy of the system.

Helium atoms in atmosphere are not catalyst but play a role in forming surface of liquid carbon. Our results of standard ab initio simulation show a clear presence of interface between liquid carbon and helium gas. Therefore, in the present simulations, instead of the presence of helium atoms, a stochastic force field acts on carbon atoms at surface region, which is formulated by the collision model of ideal gas. When liquid carbon of $1.5\pm0.05$ g/cm$^3$ is evaporated at 6000 K, for example, about one atom per nm$^2$×ps are evaporated from the liquid surface. About 70% of evaporated atoms are dimers and trimers and others are small carbon chainlike clusters with no more than 8 atoms. The collision effect is included, as the stochastic force field, also in the CNT part, since the helium atmosphere exists around CNT.

During simulations of growth process of a single CNT, we move the CNT by controlling the z-coordinate of the center of mass of the upper tip atoms. The local temperature of the upper CNT tip is kept to be lower than that of the liquid part during the simulation process. Since the number of the controlled atoms in the upper tip is about 2% of the total, their influence on the averaged temperature of the whole system can be neglected.

In the simulation of synthesis of Y junction, the stochastic force field only acts on carbon atoms at liquid surface region not at the CNT part for simplicity, since our simulation results indicate that the force field on CNT part is not crucial for the CNT growth.

Acknowledgements

This work was supported by NSFC grants 60401013. Computation was partially carried out at the Supercomputer Center, Institute for Solid State Physics, University of Tokyo.

Correspondence and requests for materials should be addressed to S. Z. (szhang@coral.t.u-tokyo.ac.jp).

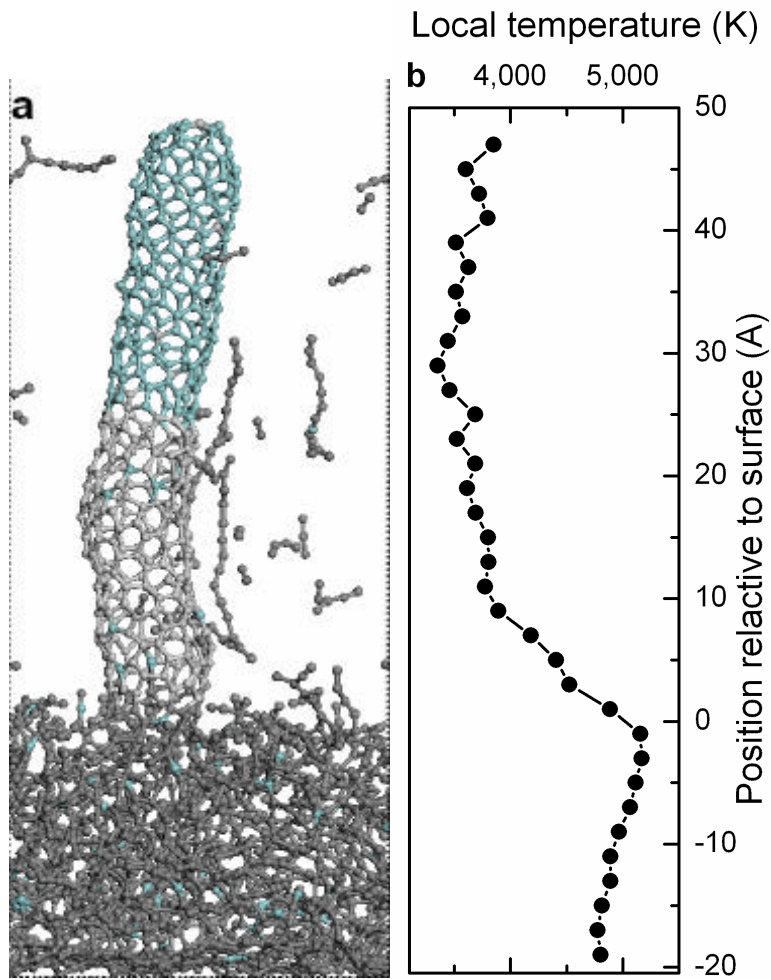

Figure 1: Simulation result of plunging CNT (9,0) into liquid carbon of 1.7 g/cm³ and drawing out with velocity of 30 m/s. **a**, Final configuration, the atoms in initial CNT are painted green. Some of them melt in liquid. Atoms of light grey are initially dispersed in the whole liquid region and then form the elongated part of the CNT. **b**, Local temperature distribution in CNT and liquid part of final configuration. The averaged temperature is plotted as a function of height.

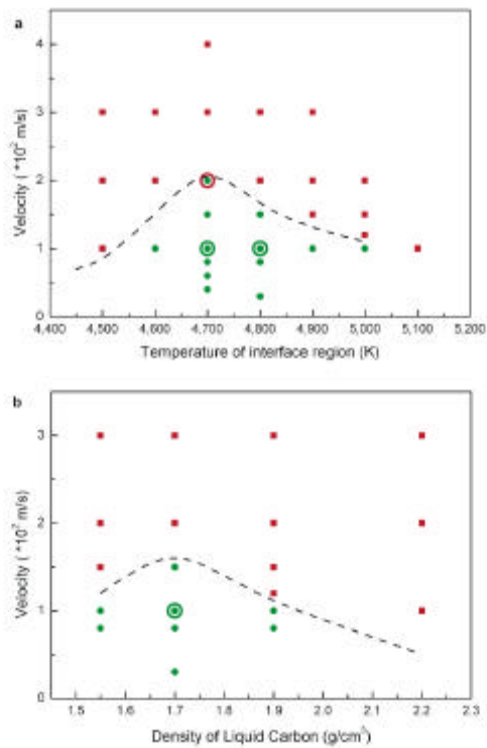

Figure 2: Summary of the simulation results of growth process of CNT (9,0). **a**, The density of liquid is 1.7±0.1 g/cm³. **b**, The average kinetic energy of total system is 4800 K. Green dots indicate the parameter set with successful growth longer than 1.5 nm. Red squares indicate the parameter set with failure, in which the growth longer than 1.5 nm is not achieved. Green circles show the parameter set where simulations are successful for three times (all trials). Red circle shows the case where two trials from three are successful and fail in another time. The dash line indicates the upper limit of growth velocity just for eye guide.

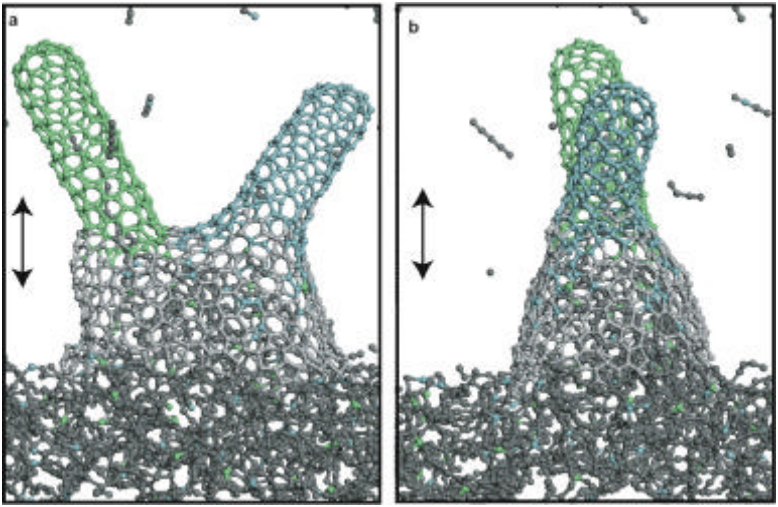

Figure 3: Fabrication of a Y junction from two (9,0) CNTs, The lower part of the Y junction is created and elongated by longer than 1.5 nm. Atoms in initial CNTs are painted green or blue. Atoms of light grey are initially dispersed in the whole liquid region and then form the lower part of the Y junction. **a**, Front view of simulation result. **b**, Side view of simulation result. Scale bar, 1 nm

Due to the limit of file size, the supplementary video files are replaced by a series of snapshots.

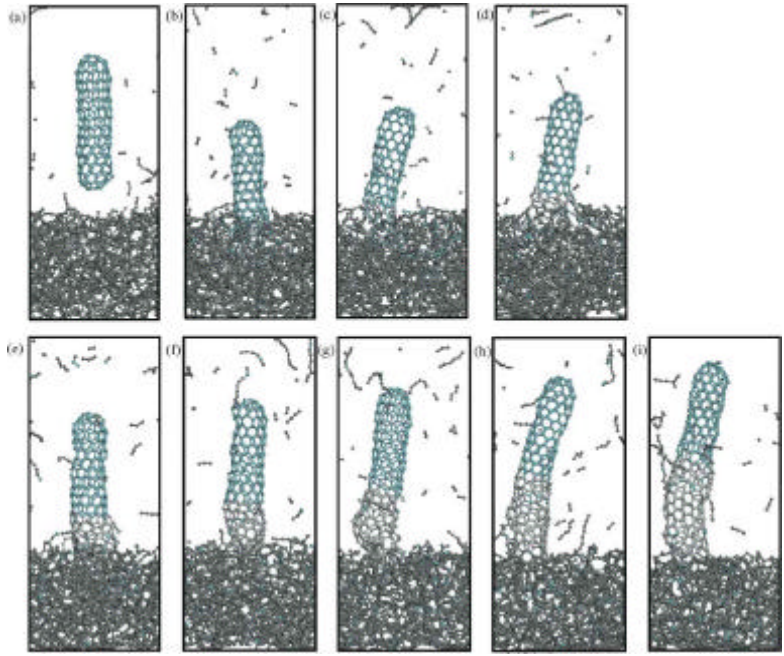

Supplementary Video 1: Growth process of a single-wall carbon nanotube from liquid carbon. As initial configuration, a (9,0) CNT is put above the liquid surface. The CNT firstly moves down with velocity of 300 m/s till it plunges into liquid about 1 nm depth, and then turns to rise with a constant velocity of 30 m/s. As a result, atoms of light grey are initially dispersed in the whole liquid region and then form the elongated part of the CNT. The CNT is elongated longer than 2 nm, and can be elongated further. The colouring in this Video is the same as that in Fig. 1a. The total simulation time is 86 ps.

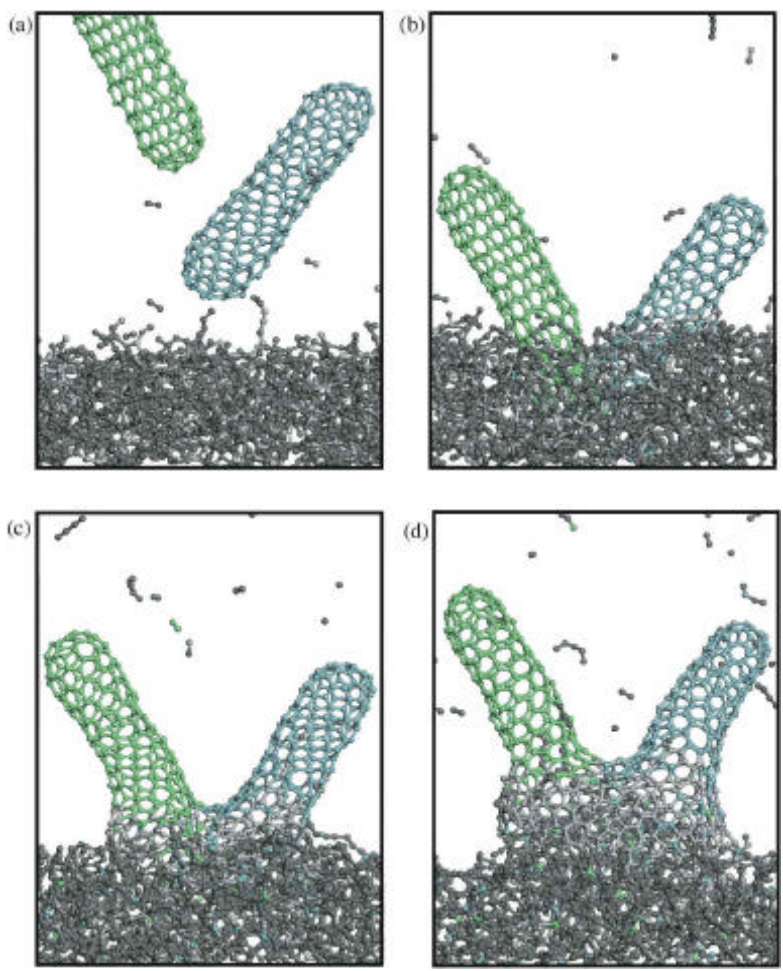

Supplementary Video 2: Fabrication of a Y junction from two single-wall CNTs. Two (9,0) CNTs fall down and plunge into liquid carbon of 1.7 g/cm3, and then rise simultaneous with the same velocity of 20 m/s. As a result, the lower part of the Y junctions is created and elongated by longer than 1.5 nm from liquid, and can be elongated further. The colouring in this Video is the same as that in Fig. 3. The total simulation time is 74 ps.